\documentclass[twocolumn,nofootinbib,aps,superscriptaddress,preprintnumbers]{revtex4}

\usepackage{graphicx}

\newif\ifpdf
\ifx\pdfoutput\undefined
\pdffalse 
\else
\pdfoutput=1 
\pdftrue
\fi

\def\OMIT#1{{}}
\def\lqcd{\ensuremath{\Lambda_{\rm QCD}}}

\def\GeV{\mbox{GeV}}

\def\d{{\rm d}}

\def\Eg{\ensuremath{E_\gamma}}
\def\Pg{\ensuremath{P_\gamma}}
\def\Pp{\ensuremath{P_+}}
\def\pphat{\ensuremath{\hat p_+}}

\def\kphat{\ensuremath{\hat k_+}}
\def\xb{\ensuremath{\bar x}}

\def\mbbar{\overline{m}_b}

\newcommand{\nn}{\nonumber}
\newcommand{\beq}{\begin{equation}}
\newcommand{\eeq}{\end{equation}}
\newcommand{\beqa}{\begin{eqnarray}}
\newcommand{\eeqa}{\end{eqnarray}}

\begin{document}
\ifpdf
\DeclareGraphicsExtensions{.pdf, .jpg}
\else
\DeclareGraphicsExtensions{.eps, .jpg}
\fi

\preprint{ \vbox{\vspace*{.75cm} \hbox{MPP--2005--1} \hbox{LBNL--56798}
\hbox{hep-ph/0502134} }}

\title{\boldmath Perturbative corrections to the determination of $V_{ub}$\\
  from the \Pp\ spectrum in $B\to X_u\ell\bar\nu$}

\vspace*{1.5cm}

\author{Andre H.\ Hoang}
\affiliation{Max-Planck-Institut f\"ur Physik, Werner-Heisenberg-Institut,
  F\"ohringer Ring 6, 80805 M\"unchen, Germany}
\author{Zoltan Ligeti}
\affiliation{Ernest Orlando Lawrence Berkeley National Laboratory,
  University of California, Berkeley, CA 94720}
\author{Michael Luke}
\affiliation{Department of Physics, University of Toronto,
  60 St.\ George Street, Toronto, Ontario, Canada M5S 1A7}

\begin{abstract}

We investigate the relation between the \Eg\ spectrum in $B\to X_s\gamma$ decay
and the \Pp\ spectrum in semileptonic $B\to X_u\ell\bar\nu$ decay (\Pp\ is the
hadronic energy minus the absolute value of the hadronic three-momentum), which
provides in principle the theoretically simplest determination of $|V_{ub}|$
from any of the ``shape function regions" of $B\to X_u\ell\bar\nu$ spectra.  We
calculate analytically the \Pp\ spectrum to order $\alpha_s^2\beta_0$, and study
its relation to the $B\to X_s\gamma$ photon spectrum to eliminate the leading
dependence on nonperturbative effects.  We compare the result of fixed order
perturbation theory to the next-to-leading log renormalization group improved
calculation, and argue that fixed order perturbation theory is likely to be a
more appropriate expansion.  Implications for the perturbative uncertainties in
the determination of $|V_{ub}|$ from the \Pp\ spectrum are discussed.

\end{abstract}

\maketitle

\section{Introduction}

The determination of the magnitude of the Cabibbo-Kobayashi-Maskawa (CKM) matrix
element $V_{ub}$ via inclusive decays is theoretically involved, because the
experimental cuts required to suppress the $B\to X_c\ell\bar\nu$ background tend
to restrict the $B\to X_u\ell\bar\nu$ phase space in a way that gives rise to
theoretical complications~\cite{Luke:2003nu}.  The local operator product
expansion (OPE)~\cite{OPE,book} breaks down in the regions $E_\ell >
(m_B^2-m_D^2)/(2m_B)$ and $m_X < m_D$ \cite{Falk:1997gj,Dikeman:1997es} (where
$E_\ell$ is the charged lepton energy and $m_X$ the hadronic invariant mass),
because numerically $m_c^2 \sim \lqcd m_b$.  In these regions an expansion in
$\lqcd/m_b$ in terms of nonlocal matrix elements is still
possible~\cite{Bauer:2001mh}, and the leading term can be measured in $B\to
X_s\gamma$ decay~\cite{Neubert:1993um}. (The local OPE is valid in the $q^2 >
(m_B-m_D)^2$ region~\cite{Bauer:2000xf}, $q^2$ being the lepton invariant mass,
but there are other issues for that cut~\cite{Luke:2003nu}.)  The kinematic
region, in which the final hadronic state has high energy $\sim m_b$ but low
invariant mass $\sim \sqrt{\lqcd m_b}$, is typically known as the ``shape
function region."

It was pointed out~\cite{Mannel:1999gs} (see
also~\cite{Bosch:2004th,Bosch:2004bt}) that separating $b\to u$ from $b\to c$
using another variable, $\Pp \equiv E_X - |\vec p_X|$, where $E_X$ and $\vec
p_X$ are the energy and three-momenta of the hadronic final state, may provide
advantages compared to $E_\ell$ or $m_X$.  At lowest order in $\lqcd/m_b$
$\d\Gamma_{B\to X_u\ell\bar\nu} / \d\Pp$ is proportional to $\d\Gamma_{B\to
X_s\gamma} / \d\Eg$ evaluated at $\Eg = (m_B-\Pp)/2$ (since \Pp\ in $B\to
X_s\gamma$ equals $m_B-2\Eg$).  Thus, to predict the $B\to X_u\ell\bar\nu$ rate
in the $\Pp < m_D^2/m_B$ region that is free from the charm background, we only
need to know the $B\to X_s\gamma$ photon spectrum for $\Eg >
(m_B^2-m_D^2)/(2m_B)$; i.e., a 330\,MeV region near the endpoint, which is
already precisely measured~\cite{Chen:2001fj,Koppenburg:2004fz,Aubert:2002pd}. 
This is also what is needed to predict the $E_\ell$ endpoint spectrum, but it is
significantly smaller than what is required to determine the $m_X$
spectrum~\cite{DeFazio:1999sv,Burrell:2003cf}.

A convenient way to express the relation between the \Pp\ spectrum in
semileptonic $b\to u$ decay and the photon energy spectrum in $B\to X_s\gamma$
decay is to relate weighted integrals of the two
spectra~\cite{Leibovich:1999xf} (see also~\cite{Bosch:2004th}).  
One can write
\beq\label{Wdef0}
\int_0^\Delta \d\Pp\, \frac{\d\Gamma_u}{\d\Pp} 
  \propto {|V_{ub}|^2 \over |V_{tb}V_{ts}^*|^2}\,
  \int_0^\Delta \d\Pg\, W(\Delta, \Pg)\, {\d\Gamma_s\over\d\Pg} \,,
\eeq
where we have defined  $\Pg \equiv m_B-2\Eg$, and at leading order in 
$\lqcd/m_b$ the weight function
$W(\Delta, \Pg)$ is calculable in perturbation theory.

In the shape function region, $\Delta\sim \lqcd$, the perturbative expansion of
$W$ contains logarithms of the ratio of scales $\sqrt{m_b \lqcd}/m_b$.  For
$m_b\gg\lqcd$, these logarithms are large and can spoil the convergence of
perturbation theory, so must be resummed using renormalization group
techniques.  This can be done using traditional perturbative QCD methods
\cite{Korchemsky:1994jb, Akhoury:1995fp, Leibovich:1999xf, Leibovich:1999xf2} 
or by using the Soft-Collinear Effective Theory (SCET) \cite{Bauer:2000ew}.
Currently $W$ can be extracted from known results to leading log (LL) and
next-to-leading log (NLL) accuracy.  However, for the true value
$\sqrt{m_b\lqcd} / m_b\sim 1/3$, it is not clear that the leading log expansion
is appropriate, and a fixed order calculation in $\alpha_s$ may give a better
approximation~to~$W$.\footnote{This is reminiscent of the resummation of logs of
$m_c/m_b\sim 1/3$ in exclusive $B\to D^*\ell\bar\nu$ decay at zero recoil, in
which the leading log calculation is a poor approximation to the one- or
two-loop results~\cite{Falk:1990yz,Czarnecki:1996gu}.}

In this paper we address this issue by calculating the order $\alpha_s^2\beta_0$
corrections to the  \Pp\ spectrum, where $\beta_0=11-2 n_f/3$ is the first
coefficient of the QCD $\beta$-function (we shall refer to this order as
BLM~\cite{Brodsky:1982gc}), from which we determine the corresponding
corrections to the weighting function $W$.  Since $\beta_0\sim 9$ is a large
number, such terms dominate the two-loop corrections to many
processes~\cite{Brodsky:1982gc,Luke:1994yc}.  We find that a considerable part
of the two-loop expression arises from terms which in the renormalization group
improved perturbation theory only occur at next-to-next-to leading log (NNLL)
order, and we argue that fixed order perturbation theory is a more appropriate
expansion.  We discuss the implications of this for the uncertainty in the
determination of $|V_{ub}|$.

\section{\boldmath The spectra at order $\alpha_s^2\beta_0$}

We first present analytical results for the parton level spectra in $B\to
X_u\ell\bar\nu$ and $B\to X_s\gamma$ to order $\alpha_s^2\beta_0$.  Our results
for the \Pp\ spectrum are new, whereas those for the photon spectrum are
collected from the existing literature for completeness.

At the parton level, the appropriate variables are
\beq
\pphat \equiv (v - q/m_b)\cdot n \,,\qquad 
  \bar x\equiv1-2 E_\gamma/m_b \,,
\eeq
for $B\to X_u\ell\bar\nu$ and for $B\to X_s\gamma$ decay respectively,  where
$n$ is a light-like four-vector in the direction of $-\vec q$, $v$ is the
four-velocity of the $B$ meson. These variables are simply related to the
experimentally observed hadronic variables by
\beqa\label{ppgdef}
\Pp &\equiv& E_X - |\vec p_X| =
  (m_B v - q) \cdot n = m_b \pphat + \Lambda \,, \nn\\
\Pg &\equiv& m_B-2E_\gamma = m_b\bar x + \Lambda \,,
\eeqa
where $\Lambda \equiv  m_B-m_b$.

\subsection{\boldmath The $\pphat$ spectrum in $B\to X_u\ell\bar\nu$}

The $\pphat$ spectrum to order $\alpha_s$ is given by \cite{Mannel:1999gs}
\begin{widetext}
\beqa\label{pp1}
{1\over\Gamma_0}\, {\d\Gamma_u\over \d\pphat} =
 \delta(\pphat) &-& {\alpha_s(m_b) C_F\over 4\pi} \bigg[\,
  4\, \bigg({\ln\pphat\over\pphat}\bigg)_* 
  + \frac{26}3\, \bigg({1\over\pphat}\bigg)_* 
  + \bigg(\frac{13}{36}+2\pi^2\bigg)\, \delta(\pphat) \nn\\
&+& 4\pphat^2(3-2\pphat) \ln^2\pphat 
  + \frac23\, (2-23\pphat-9\pphat^2+8\pphat^3) \ln\pphat \nn\\
&-& {316+407\pphat-1101\pphat^2+708\pphat^3-200\pphat^4+33\pphat^5-7\pphat^6
  \over 18} \,\bigg] + {\cal O}(\alpha_s^2)\,,
\eeqa
where $\Gamma_0 = G_F^2 |V_{ub}|^2\, m_b^5/ (192\pi^3)$, $C_F = 4/3$, and $m_b$
is the $b$ quark pole mass. The $*$ distributions (for $n\geq0$ integers) are
defined by 
\beq
\int_0^z \d x\, f(x)\, \bigg({\ln^n x\over x}\bigg)_* 
  = f(0)\, {\ln^{n+1} z\over n+1} 
  + \int_0^z \d x\, [f(x)-f(0)]\, {\ln^n x\over x}\,.
\eeq
Integrating Eq.~(\ref{pp1}) over $0<\pphat<1$ reproduces the total rate $1 -
(\alpha_s C_F/(4\pi))(2\pi^2-25/2)$.

The BLM part of the two-loop result may be obtained from the one-loop
calculation with an arbitrary gluon mass using the method of
Ref.~\cite{Smith:1994id}.  We calculated the spectrum for $\pphat \neq 0$, for
which only bremsstrahlung graphs contribute, and then determined the coefficient
of $\delta(\pphat)$ by comparing with the total rate~\cite{vanRitbergen:1999gs}.
The result is
\beqa\label{pp2}
{1\over\Gamma_0} {\d\Gamma_u^{\rm (BLM)}\over \d\pphat}
&=& {\alpha_s^2(m_b)\over \pi^2}\, \beta_0\, \Bigg\{
  \frac12\, \bigg({\ln^2\pphat\over\pphat}\bigg)_* +
  \frac12\, \bigg({\ln\pphat\over\pphat}\bigg)_* +
  \bigg({\pi^2\over18}-{113\over 72}\bigg) \bigg({1\over\pphat}\bigg)_* -
  \bigg(\frac43\, \zeta_3 + \frac{41}{72}\pi^2 - \frac{1333}{1728}
  \bigg)\, \delta(\pphat) \nn\\*
&-& \frac{2 \pphat^2(3 - 2\pphat)}{3}\, \Big[
  {\rm Li}_3(\pphat) + {\rm Li}_3(\pphat(2 - \pphat))
  - 2 {\rm Li}_3(1/(2 - \pphat)) + \frac14 {\rm Li}_3((1 - \pphat)^2)\nn\\*
&&{} \quad - {\rm Li}_2(\pphat) \ln(\pphat(2-\pphat)^2)
  - \frac{\pi^2}6 \ln(\pphat(2 - \pphat))
  + \frac12 \ln(1 - \pphat) \ln^2 \pphat - \frac56 \ln^3 \pphat
  + \frac13 \ln^3(2 - \pphat) \Big] \nn\\*
&+& \frac{6 + 11 \pphat - 32 \pphat^2 + 18 \pphat^3 - 4 \pphat^4}{18\pphat}\,
  {\rm Li}_2((1-\pphat)^2) - {\pi^2\over 18\pphat} 
  + (1-\pphat)\, {\rm Li}_2(\pphat-1) \nn\\*
&+& \frac{53}{216 (1-\pphat)}
  - \frac{674 - 1333\pphat+606\pphat^2}{216(1-\pphat)^3}\, \ln(\pphat(2-\pphat))
  + \frac{6 - 69 \pphat - 123 \pphat^2 + 100\pphat^3}{36}\, \ln^2\pphat \nn\\*
&+& \frac{605 + 184 \pphat - 237 \pphat^2 - 106\pphat^3}{108} \ln\pphat 
+ \frac{2374 + 6219 \pphat - 15589 \pphat^2 + 9890 \pphat^3 - 2352\pphat^4
  + 417 \pphat^5 - 87 \pphat^6}{864} \nn\\*
&+& \frac{530 + 137 \pphat - 1341 \pphat^2 + 840 \pphat^3 - 200 \pphat^4
  + 33 \pphat^5 - 7 \pphat^6}{216}\, \ln \frac{2-\pphat}{\pphat^2} 
  \Bigg\} ,
\eeqa
\end{widetext}
where ${\rm Li}_2(z) = -\int_0^z \d t \ln(1-t)/t$ is the dilogarithm, and ${\rm
Li}_3(z) = \int_0^z \d t\, {\rm Li}_2(t)/t$.

\subsection{\boldmath The \xb\ spectrum in $B\to X_s\gamma$}

We concentrate on the part of the spectrum that arises from the operator $O_7 =
(e/16\pi^2)\, m_b\, \bar s_L\, \sigma^{\mu\nu} F_{\mu\nu}\, b_R$.  This gives
rise to the dominant part of the photon spectrum, and while other operators also
influence the spectrum, the photon spectrum is only known analytically to order
$\alpha_s^2\beta_0$ for this piece (and for $O_8$, but not for the terms
involving the four-quark operators).  To study the convergence of the
perturbative expansions and assess the theoretical uncertainties in the \Pp\
spectrum, it is sufficient to consider $O_7$.  However, ultimately, for the
actual determination of $|V_{ub}|$ the other contributions should also be
included.

The photon spectrum to order $\alpha_s$ is given by
\cite{Ali:1990tj,Kapustin:1995nr}
\begin{widetext}
\beq\label{spect77}
{1\over\Gamma_\gamma}\, {\d\Gamma_{77}\over\d \xb} = \delta(\xb)
- {\alpha_s(m_b) C_F\over4\pi} \bigg[\,
  4 \bigg({\ln\xb\over\xb}\bigg)_* + 7 \bigg({1\over\xb}\bigg)_*
  + \bigg(5 + \frac43\,\pi^2 \bigg)\, \delta(\xb) 
  -6-3\xb+2\xb^2+2(2-\xb)\ln\xb \bigg] + {\cal O}(\alpha_s^2) ,
\eeq
where $\Gamma_\gamma = G_F^2\, |V_{tb}V_{ts}^*|^2\, \alpha_{\rm em}\, m_b^3\,
[\mbbar(m_b) C_7^{\rm eff}(m_b)]^2 / (32\pi^4)$, and we have set $\mu=m_b$ for
convenience. The BLM correction to the photon spectrum due to $O_7$ may be
obtained by combining results given in~\cite{Bieri:2003ue}
and~\cite{Ligeti:1999ea}; again, setting $\mu=m_b$ it reads
\beqa\label{spect77blm}
{1\over\Gamma_\gamma}\, {\d\Gamma_{77}^{\rm (BLM)}\over\d \xb} 
&=& {\alpha_s^2(m_b)\over \pi^2}\, \beta_0\,
  \bigg[ \frac12\, \bigg({\ln^2\xb\over\xb}\bigg)_*
  + \frac{13}{36}\, \bigg({\ln\xb \over\xb}\bigg)_*
  + \bigg(\frac{\pi^2}{18}-\frac{85}{72}\bigg)\, \bigg({1\over\xb}\bigg)_* 
  - \bigg(\frac13\,\zeta_3 + \frac{91}{216}\pi^2+\frac{631}{432}\bigg)\, \delta(\xb) \\*
&&{} + \frac{(6+6\xb-3\xb^2){\rm Li}_2(1-\xb) - \pi^2}{18\xb} + \frac{2-\xb}4\, \ln^2\xb
  - \frac{38-33\xb+7\xb^2+6\xb^3}{36(1-\xb)}\, \ln\xb 
  + \frac{66+21\xb-38\xb^2}{72} \bigg] . \nn
\eeqa
\end{widetext}

\section{Relation between the spectra}

One can define a weighting function, $W$, that relates, in the shape function
regions, weighted integrals of the photon energy spectrum in $B\to X_s\gamma$ to
integrals of the \Pp\ spectrum in $B\to X_u l\nu$:
\beqa\label{Wdef}
\int_0^\Delta \d\Pp\, \frac{\d\Gamma_u}{\d\Pp} 
&=& {|V_{ub}|^2 \over |V_{tb}V_{ts}^*|^2}\,
  {\pi \over 6\, \alpha_{\rm em}\, C_7^{\rm eff}(m_b)^2}\,
  {m_B^2\over \mbbar(m_b)^2} \nn\\*
&\times& \int_0^\Delta \d\Pg\, W(\Delta, \Pg)\, {\d\Gamma_{77}\over\d\Pg} \,.
\eeqa
The weighting function $W$  can be computed perturbatively at leading order in
$\lqcd/m_b$, because the shape function contribution that captures the leading
nonperturbative physics at scales of order \lqcd\ is identical in the two
spectra.  The corresponding perturbation series depends on the scales $m_b$ and
$\sqrt{m_b\Lambda_{\rm QCD}}$.  The function $W$ is also free of logarithms of
the form $\alpha_s^n \ln^{m}(m_b^2/\mu^2)$, $m=n+1, \dots, 2n$, which are
universal in the two spectra and cancel from the relation in Eq.~(\ref{Wdef}).

The origin of the $m_B^2/\mbbar^2$ factor in the definition of $W$ deserves
comment. 
At leading order in $\lqcd/m_b$, the effects of the shape function
$f(\omega)$ may be simply included by making the replacement
\beq\label{mbsmear}
m_b\to m_b^* = m_b+\omega\,, 
\eeq
in the tree level partonic rate, and then convoluting the differential rate with
the shape function $f(\omega)$~\cite{Neubert:1993um},
\beq\label{smearing}
\d\Gamma = \int \d\Gamma^{\rm parton}\big|_{m_b\to m_b^*}\, f(\omega)\, 
  \d\omega.
\eeq
This prescription also generalizes to higher orders in $\lqcd/m_b$, provided
that the replacement (\ref{mbsmear}) is only applied to factors of $m_b$ which
arise from kinematics, and not from coefficients of operators in the Hamiltonian
\cite{Burrell:2003cf} and correctly reproduces the class of ``kinematic"
subleading effects (proportional to the leading order shape function) which do
not arise from the expansion of the heavy quark fields.

Applying this procedure to the tree level spectra, we have
\beqa
{\d\Gamma_u\over \d\Pp} &\propto& \int(m_b+\omega)^5\,
  \delta(\Pp-\Lambda+\omega)\, f(\omega)\, \d\omega \nn\\*
&=& \int(m_B-\tilde\omega)^5\,
  \delta(\Pp-\tilde\omega)\, f(\Lambda-\tilde\omega)\, \d\tilde\omega 
\eeqa
for semileptonic decays (where $\tilde\omega\equiv\Lambda-\omega$), and
\beq
{\d\Gamma_{77}\over \d\Pg} \propto \mbbar(m_b)^2 
  \int(m_B-\tilde\omega)^3\, \delta(\Pg-\tilde\omega)\, 
  f(\Lambda-\tilde\omega)\, \d\tilde\omega  \nn\\
\eeq
for radiative decays, where two powers of $\mbbar(m_b)$ originate from $C_7$ in
the effective Hamiltonian to which  the replacement (\ref{mbsmear}) does not
apply.  This gives
\beqa\label{kin1m}
\int_0^\Delta {\d\Gamma_u\over \d\Pp}\, \d\Pp &\propto& \int_0^\Delta 
  {(m_B-\Pg)^2\over\mbbar(m_b)^2}\, {\d\Gamma_{77}\over \d\Pg}\, \d\Pg \\*
&&\hskip-0.5in = {m_B^2\over\mbbar(m_b)^2}\int_0^\Delta 
   \left(1-{2\Pg\over m_B}+\dots\right) {\d\Gamma_{77}\over \d\Pg}\, \d\Pg .\nn
\eeqa
Thus, it is natural to pull the leading factor of $m_B^2/\mbbar(m_b)^2$ out of
the definition of $W$.  As we will show, factoring out this term rather than
leaving the ``partonic" factor $m_b^2/\mbbar(m_b)^2$ in the definition of $W$
dramatically reduces the absolute size of the perturbative corrections to $W$.

To calculate $W$, we first expand both parton level spectra in powers of
$\pphat$ and $\xb$ respectively, since in the shape function region both are
${\cal O}(\lqcd/m_b)$.  At leading order in $\lqcd/m_b$, this simply corresponds
to keeping the first four terms in Eqs.~(\ref{pp2}) and (\ref{spect77blm}).  At
leading order in the $\lqcd/m_b$ expansion relevant for the shape function
regime, the Feynman diagrams which give the coefficient of the shape function
give the partonic result with the substitution $\pphat\to\pphat+\kphat$ and
$\xb\to\xb+\kphat$, where $\kphat$ is the light-cone component of the residual
momentum of the heavy quark, which is of the same order.  (The dimensionless
variables $\kphat$ and $\hat\omega$ have support between $-\infty$ and
$\Lambda/m_b$.)  Thus, we have
\begin{widetext}
\beqa\label{pp2exp}
{1\over\Gamma_0}\, {\d\Gamma_u^{\rm (BLM)}\over \d\pphat}
&=& {\alpha_s^2(m_b)\over \pi^2}\, \beta_0\, \delta(\hat\omega-\kphat)\,
  \bigg[ \frac12 \bigg({\ln^2(\pphat+\hat\omega)\over\pphat+\hat\omega}\bigg)_*
  + \frac12 \bigg({\ln(\pphat+\hat\omega)\over\pphat+\hat\omega}\bigg)_* 
  + \bigg({\pi^2\over18}-{113\over 72}\bigg) 
  \bigg({1\over\pphat+\hat\omega}\bigg)_* \nn\\*
&&\qquad\qquad\qquad\qquad\quad{}\ 
  - \bigg(\frac43\, \zeta_3 + \frac{41}{72}\pi^2 - \frac{1333}{1728}
  \bigg)\, \delta(\pphat+\hat\omega)+\dots \bigg] ,
\eeqa
and similarly for $\d\Gamma_{77}/\d \xb$.  This gives, for each spectrum, the
matching conditions onto the leading nonlocal operator $O_0(\omega)
= \bar b\, \delta(\omega-iD_+)\, b$ in the OPE.  The $B$ meson matrix element of
$O_0(\omega)$ gives the leading order $b$ quark light-cone distribution
function, or shape function.  In principle, to determine the coefficient of
$O_0(\omega)$ we must include the radiative corrections to the parton level
matrix element of $O_0(\omega)$; however, since these terms are common to both
spectra and therefore drop out of $W$, we do not need to worry about them here.

Including the full one-loop corrections and the BLM two-loop contributions, and
setting the scale $\mu=m_b$ for simplicity, the function $W$ is given by
\beqa\label{Wlo}
W(\Delta, \Pg) &=& 1 + \frac{C_F \alpha_s(m_b)}{4\pi}\, \bigg(
  \frac53 \ln{m_b\over \Delta-\Pg}
  - \frac{2\pi^2}{3} + \frac{167}{36} \bigg) \nn\\*
&&\ \, {} + \frac{C_F \alpha_s^2(m_b)}{(4\pi)^2}\, \beta_0
  \bigg( \frac56 \ln^2{m_b\over \Delta-\Pg}
  + \frac{14}{3} \ln{m_b\over \Delta-\Pg} - \frac{16\pi^2}{9} 
  + \frac{3857}{144} - 12\zeta_3 \bigg)+\dots  
\eeqa
where the ellipses denote non-BLM two-loop terms, higher orders in perturbation
theory and nonperturbative corrections suppressed by powers of $\lqcd/m_b$.
Eq.~(\ref{Wlo}) is the main result of this paper.

It is instructive to compare Eq.~(\ref{Wlo}) with the next-to-leading log result
in SCET.  Using the results of Refs.~\cite{Bosch:2004th, Neubert:2004dd} 
(see also~\cite{Bauer:2003pi}), we find
\beq\label{NLLresum}
W^{\rm NLL}(\Delta, \Pg) = T(a) \left\{ 1 + {C_F\alpha_s(m_b)\over 4\pi}\, H(a) 
  + {C_F\alpha_s(\mu_i)\over 4\pi} 
  \left[ 4 f_2(a)\ln{m_b(\Delta-\Pg)\over\mu_i^2}
  - 3 f_2(a) + 2 f_3(a)\right] \right\} ,
\eeq
where
\beqa\label{NLLcoeffs}
T(a) &=& {2(6-a)\over(4-a)(3-a)}\,, \qquad\qquad\qquad
H(a) = - {4(486-389 a+103 a^2-9 a^3)\over (6-a)(4-a)^2(3-a)^2}
  - 4 \psi^\prime(3-a) - c\, f_2(a)\,, \nn\\
f_2(a) &=& -{30-12 a+a^2\over (6-a)(4-a)(3-a)}\,, \qquad \,
f_3(a) = {2(138-90 a+18 a^2-a^3)\over(6-a)(4-a)^2(3-a)^2} \,,
\eeqa
$\psi'$ is the derivative of the digamma function, $\psi(z) = \Gamma'(z) /
\Gamma(z)$, and
\beq
a = {\Gamma_0\over \beta_0}\, \ln{\alpha_s(\mu_i)\over \alpha_s(m_b)} \,, \qquad
c = {4\over\beta_0} \left({\Gamma_1\over\Gamma_0}-{\beta_1\over\beta_0}\right)
  \left( {\alpha_s(\mu_i)\over\alpha_s(m_b)} - 1 \right) .
\eeq
Here $\Gamma_0 = 4 C_F$ and  $\Gamma_1 = 8 C_F (67/6 - \pi^2/2 - 5n_f/9)$ are
the first two coefficients of the cusp anomalous dimension, and $\beta_1 =
102-38 n_f/3$. Setting $\mu_i \sim \lqcd m_b$ sums all leading and
subleading logarithms of the form $\alpha_s^n \ln^n(\mu_i/m_b)$ and $\alpha_s^n
\ln^{n-1}(\mu_i/m_b)$.  Expanding Eq.~(\ref{NLLresum}) to order
$\alpha_s^2(m_b)$, we find
\beqa\label{Wexpanded1}
W^{\rm NLL}(\Delta, \Pg) = 1 
&+& \frac{C_F\alpha_s(m_b)}{4\pi}\, \bigg(\frac53 
\ln{m_b^2\over\mu_i^2}\bigg)
  + \frac{C_F\alpha_s^2(m_b)}{(4\pi)^2}\,
  \bigg(\frac{5\beta_0}6 + \frac{92}{27}\bigg) \ln^2{m_b^2\over \mu_i^2} \nn\\*
&+& \frac{C_F\alpha_s(m_b)}{4\pi}\, \bigg( 
  \frac53 \ln{\mu_i^2\over m_b(\Delta-\Pg)} 
  - \frac{2\pi^2}{3} + \frac{167}{36} \bigg) \\*
&+& \frac{C_F\alpha_s^2(m_b)}{(4\pi)^2}\, \bigg[
  \bigg(\frac{5\beta_0}3 + \frac{184}{27}\bigg)
  \ln{\mu_i^2\over m_b(\Delta-\Pg)} 
  + \frac{14\beta_0}3 + \frac{64}3 \psi^{\prime\prime}(3) 
  - {85\pi^2\over 27} + {1234\over 81} \bigg] \ln{m_b^2\over\mu_i^2} 
  + \ldots\,, \nn
\eeqa
where the first line is the leading log result, and the second and third lines
are subleading logs.  Note that the renormalization group equations sum logs of
$m_b^2/\mu_i^2$, not $m_b(\Delta-\Pg)/\mu_i^2$, since for $\mu_i^2\sim\lqcd m_b$
only the first are parametrically enhanced.  The result is formally independent
of $\mu_i$, as can be seen by re-expanding the logs in
Eq.~(\ref{Wexpanded1}),
\beqa\label{Wexpanded2}
W^{\rm NLL}(\Delta, \Pg) = 1 
&+& \frac{C_F \alpha_s(m_b)}{4\pi}\, \bigg( \frac53 \ln{m_b\over \Delta-\Pg}
  - \frac{2\pi^2}{3} + \frac{167}{36} \bigg) 
+ \frac{C_F \alpha_s^2(m_b)}{(4\pi)^2}\, \bigg[
\left(\frac{5\beta_0}6+\frac{92}{27}\right)\ln^2{m_b\over \Delta-\Pg}\nn\\*
&+&\left(\frac{14\beta_0}3 + \frac{64}3 \psi^{\prime\prime}(3) 
  - {85\pi^2\over 27} + {1234\over 81}\right) \ln{m_b\over \Delta-\Pg}
  + \dots \bigg] + \dots \,,
\eeqa
where we have dropped terms of order $\alpha_s^2
\ln^n[m_b(\Delta-\Pg)/\mu_i^2]$, which are next-to-next-to-leading order.  This
result agrees with the corresponding one-loop and two-loop BLM terms in
Eq.~(\ref{Wlo}).

Equations (\ref{Wlo}) and (\ref{Wexpanded2}) both provide approximations to
the full expression for $W$ at two-loops, so it is instructive to compare them. 
Numerically, the ${\cal O}(\alpha_s^2)$ terms are
\beqa\label{W2loop}
W^{(\alpha_s^2)} &=& {C_F\alpha_s^2(m_b)\over(4\pi)^2} 
  \bigg[ (0.83\beta_0 + 3.41) \ln^2{m_b\over \Delta-\Pg} 
  + (4.67\beta_0 - 19.1) \ln{m_b\over \Delta-\Pg} - (5.19\beta_0+c_0)\bigg] 
  \nn\\*
&=& {C_F\alpha_s^2(m_b)\over(4\pi)^2} 
  \bigg[ (6.94{\beta_0\over 25/3} + 3.41) \ln^2{m_b\over \Delta-\Pg} 
  + (38.9{\beta_0\over 25/3} - 19.1) \ln{m_b\over \Delta-\Pg} 
  - (43.2{\beta_0\over 25/3}+c_0)\bigg],
\eeqa
\end{widetext}
where a complete two-loop calculation is required to determine $c_0$. 

For both the double and single log terms, the BLM-enhanced term is about a
factor of two larger than the non-BLM term, which suggests that it may also
dominate the nonlogarithmic term to a similar degree.   In contrast, the leading
log approximation is clearly poorly behaved at this order:  for
$\mu_i^2/m_b^2\sim(\Delta-\Pg)/m_b=1/9$, the double, single and  non-logarithmic
terms in Eq.~(\ref{W2loop}) are in the ratio
\beq
{\cal O}(\log^2) : {\cal O}(\log) : {\cal O}(\log^0) 
  = 1 : 0.87 : (-0.86-0.02 c_0) \,.
\eeq
This reflects the fact that the logarithmic enhancement is not sufficient for
the double log to dominate over the single or zero log terms, nor for the single
log to dominate over the BLM-enhanced piece of the zero log term.  The same
conclusion is reached by comparing the size of the LL and NLL terms in
Eq.~(\ref{Wexpanded1}).  This suggests that a fixed order calculation, rather
than a leading log calculation, is more appropriate for $W$.  We will discuss
the numerical significance of this for the extraction of $|V_{ub}|$, along with
the poor convergence of both the leading log and fixed order calculations, in
the next section.

Another source of corrections to $W$ comes from terms suppressed by
$\lqcd/m_b$.  Some of these modify $W$ without introducing new unknown hadronic
matrix elements, while  others involve nonperturbative matrix elements of
higher-dimension nonlocal operators~\cite{Bauer:2001mh}, which cannot presently
be computed.  The mismatch of the powers of $m_b$ in the two spectra gives rise
to the order $\lqcd/m_b$ correction to $W$ contained in Eq.~(\ref{kin1m}),
$-2\Pg / m_B$.  At the same order, there are also corrections that come from
expanding the $b$ quark fields in powers of $\lqcd/m_b$.  These effects are
sensitive the Dirac structure of the current, so do not cancel from $W$.  We can
extract the full $\lqcd/m_b$ correction from Ref.~\cite{Lee:2004ja}, and find
\beq\label{W1m}
W^{(\lqcd/m)} = -{8\Pg-2\Lambda \over 3m_B} + \ldots \,.
\eeq
The corrections that depend on subleading shape functions are also given
in~\cite{Lee:2004ja} and involve five unknown nonperturbative functions, whose
effects we do not attempt to model here.  Some of the ${\cal O}(\alpha_s)$
corrections to Eq.~(\ref{W1m}) can be obtained by expanding the results in
Sec.~II to higher order in $\pphat$ and $\xb$.  We find that these are small
compared to the perturbative uncertainties in $W$ at leading order in
$\lqcd/m_b$.

\section{Implications and discussions}

\begin{figure*}[t]
\centerline{\includegraphics*[width=1.25\columnwidth]{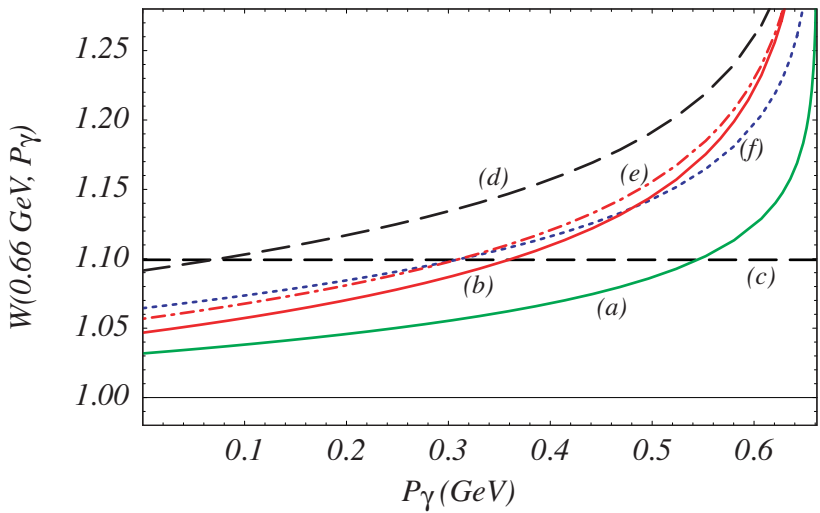}}
\caption{$W(\Delta, \Pg)$ as a function of $\Pg$.  At tree level $W = 1$, and
the curves include different orders in the expansion: 
(a) the ${\cal O}(\alpha_s)$ result;
(b) all known ${\cal O}(\alpha_s^2)$ terms, Eq.~(\ref{W2loop}) with $c_0=0$;
(c) the LL RGE resummed result; 
(d) the NLL RGE resummed result, Eq.~(\ref{NLLresum});  
(e) the ${\cal O}(\alpha_s^2\beta_0)$ result, Eq.~(\ref{Wlo}) and
(f) the NLL expression expanded to ${\cal O}(\alpha_s^2)$,
Eq.~(\ref{Wexpanded1}).}
\label{fig:Wplots}
\end{figure*}

To eliminate the charm background kinematically from semileptonic $B\to
X_u\ell\bar\nu$ decays, one has to impose a cut $\Pp < m_D^2/m_B \simeq
0.66\,\GeV$.  Then measurements of the \Pp\ spectrum in $B\to X_u\ell\bar\nu$
and the \Pg\ spectrum in $B\to X_s\gamma$, together with the theoretical input
of $W(\Delta, \Pg)$ will allow a determination of $|V_{ub}|$ using
Eq.~(\ref{Wdef}).

In Fig.~\ref{fig:Wplots} we plot $W(\Delta, \Pg)$ for $\Delta = 0.66\,$GeV,
$\alpha_s(m_b) = 0.22$ and $m_b = 4.8\,$GeV in different approximations.  At
tree level, $W(\Delta, \Pg) = 1$.  Curve (a) is the order $\alpha_s$ result,
while the result with all known $\alpha_s^2$ contributions is shown in curve
(b)  [i.e., Eq.~(\ref{W2loop}), the BLM result plus the non-BLM coefficient of
the double and single logs]. The dashed black curves show the RGE resummed
results for $\mu_i = \sqrt{m_b\, \Delta} \simeq 1.78\,$GeV.  The straight line
(c) is the LL result, whereas the long dashed curve (d) is the NLL result in
Eq.~(\ref{NLLresum}).  The LL and NLL results use 1- and 2-loop running for
$\alpha_s$, respectively.   Finally, the dot-dashed curve (e) contains just the
BLM terms at order $\alpha_s^2$, Eq.~(\ref{Wlo}), while the short dashed curve
(f) contains NLL expression expanded to order $\alpha_s^2$,
Eq.~(\ref{Wexpanded1}).

The difference between the dashed (d) and (f) curves shows that the terms in the
NLL sum beyond ${\cal O}(\alpha_s^2)$ are not negligible.  However, the logs
that are summed do not dominate over other contributions in the perturbation
series; e.g., the difference between Eqs.~(\ref{Wexpanded1}) and
(\ref{Wexpanded2}) is NNLL, but it is comparable to the non-BLM LL and NLL terms
at ${\cal O}(\alpha_s^2)$.  Therefore, we view the fixed order result, curve
(b), as our best estimate of the perturbation theory prediction of $W$.  The
difference of these curves provides an estimate of the uncertainty related to
higher order uncalculated terms.  In addition, the $\mu_i$-dependence of the NLL
resummed result is larger using two-loop than one-loop running for $\alpha_s$,
also indicating that renormalization group improved perturbation theory may not
lead to an improved expansion.

\begin{table}[t]
\caption{Weighted integral $\int_0^\Delta \d\Pg\, W(\Delta, \Pg)\,
(\d\Gamma_s/\d\Pg)$, normalized to $\int_0^\Delta \d\Pg\, (\d\Gamma_s/\d\Pg)$
for $\Delta=0.66\,\GeV$, taking the simple parametrization (\ref{dgexp}) of the
experimental photon energy spectrum.}
\label{finaltable}
\vspace*{2pt}
\begin{tabular}{c||c|c||c|c||c} \hline\hline
Tree~  &  ~${\cal O}(\alpha_s)$~  &  ~${\cal O}(\alpha_s^2\beta_0)$~  
  &  ~~LL~~  &  ~NLL~  &  ~all known ${\cal O}(\alpha_s^2)$ \\ \hline
1&  1.10  &  1.19   & 1.10  &  1.22  &  1.18  \\ \hline\hline
\end{tabular}
\end{table}

To assess the significance of these results for the determination of $|V_{ub}|$
using Eq.~(\ref{Wdef}), we integrated these various expansions of
$W(\Delta, \Pg)$ against a simple parametrization of the experimentally measured
$B\to X_s\gamma$ spectrum \cite{Koppenburg:2004fz}.  For the optimal cut,
$\Delta = 0.66\,$GeV, the integral on the right-hand side of Eq.~(\ref{Wdef})
normalized to that integral at tree level (corresponding to $W=1$) is shown in
Table \ref{finaltable} at order $\alpha_s$, $\alpha_s^2\beta_0$, using the LL
and NLL RGE resummations, and all known ${\cal O}(\alpha_s^2)$ terms.  The
simple parametrization 
\beq\label{dgexp}
{\d\Gamma_s\over \d\Pg}\bigg\vert_{\rm exp.}
  \propto x^{(a-1)} e^{-a x}, \qquad
x \equiv {\Pg\over\widetilde\Lambda},
\eeq
with $\widetilde\Lambda\sim 0.9\,{\rm GeV}$ and $a\sim 6.1$ provides a crude
but, for our purposes, sufficient fit to the data,\footnote{A parameterization
of the physical $B\to X_s\gamma$ spectrum is not available, only the raw
spectrum or fits to shape function parameters~\cite{Limosani:2004jk}.  However,
the shape function fit includes ${\cal O}(\alpha_s)$ corrections differently
than how they enter $W$.   Thus, the numbers in Table \ref{finaltable} should be
taken only as indicative of the size of the corrections.} as the ratios in Table
\ref{finaltable} are quite insensitive to the precise shape of the spectrum. 
(Using the parameters $\widetilde\Lambda\sim 0.66\,{\rm GeV}$ and $a\sim 3.3$,
which is the shape function fit rather than the photon spectrum
\cite{Limosani:2004jk}, and so corresponds to a rather different shape, only
changes the entries in Table~I to 1, 1.08, 1.15, 1.10, 1.18, 1.14,
respectively.  However, it changes $\int_0^\Delta \d\Pg\, (\d\Gamma_s/\d \Pg)$
by about a factor of two.)

The convergence of the result in Table \ref{finaltable} going from tree level to
${\cal O}(\alpha_s)$ to ${\cal O}(\alpha_s^2\beta_0)$ is poor, and that going
from tree level to LL to NLL resummation is worse.  This poor behavior of the
perturbation series may be related to the fact that in the nonlocal OPE there
are ${\cal O}(\lqcd/m_b)$ nonperturbative corrections to Eq.~(\ref{Wdef})
arising both from the explicit factor of $\Lambda$ in Eq.~(\ref{W1m}) as well as
the subleading shape functions, and so there is a renormalon ambiguity
in the perturbation series for $W$ at this order.

The subleading twist terms in Eq.~(\ref{W1m}) give a $-0.19$ correction to the
numbers in Table~1.  (The ``trivial" part of this correction, $-2\Pg/m_B$, that
is due to the mismatch of the two powers of $m_b$ smeared in the two spectra,
gives numerically the same result.)  While this is a calculable effect and not
an uncertainty, there are incalculable subleading shape functions that enter at
the same order, which can only be modelled.  Thus, the perturbative and
nonperturbative uncertainties in this determination of $|V_{ub}|$ are probably
comparable.

Because the experimental photon spectrum is peaked around $\Pg\sim 0.8\,\GeV$,
the weighted integral (\ref{Wdef}) is dominated by small values of $\Delta-\Pg$,
increasing the importance of the logarithmically enhanced terms in
(\ref{Wexpanded2}).  However, this numerically large logarithm is not summed by
the RGE, since it is not logarithms of $(\Delta-\Pg)/m_b$ but rather of
$\mu_i/m_b$ which are summed.

Had we not factored out $m_B^2/\mbbar^2$ in the definition of $W$ in
Eq.~(\ref{Wdef}), there would be an additional perturbative factor of
$m_b^2/\mbbar^2$ in $W$ which, when expanded out, would modify the expressions
in Eqs.~(\ref{Wlo}), $H(a)$ in (\ref{NLLcoeffs}), (\ref{Wexpanded1}),
(\ref{Wexpanded2}), (\ref{W2loop}), and (\ref{W1m}).  If we define a weighting
function $W'(\Delta, \Pg)$ in this way, then instead of Eq.~(\ref{W2loop}) we
obtain
\beqa\label{W2loopp}
W^{\prime(\alpha_s^2)} &=& {C_F\alpha_s^2(m_b)\over(4\pi)^2} 
  \bigg[ (0.83\beta_0+3.41) \ln^2{m_b\over \Delta-\Pg} \\*
&&{} + (4.67\beta_0-1.35) \ln{m_b\over \Delta-\Pg} 
  + (32.3\beta_0+c'_0)\bigg] , \nn
\eeqa
and the non-logarithmic BLM term is considerably larger than either of the
logarithmic terms.  In this case the numerical results in Table~I would read 1,
1.29, 1.51, 1.10, 1.43, 1.53, respectively, and we would have to assign a much
larger perturbative uncertainty to $W'$ than to $W$.  However, this is due to
the bad perturbative behavior of $m_b^2/\mbbar^2$, and not of the spectra
themselves.  In this case the analog of Eq.~(\ref{W1m}), $W^{\prime(\lqcd/m)} =
-8(\Pg-\Lambda)/(3m_B) + \ldots$, gives a $-0.01$ correction to the figures in
the previous sentence.  It is interesting to check the consistency of the
results.  Combining all known order $\alpha_s^2$ and $\lqcd/m_b$ terms, we
obtain
\beq\label{dum}
m_B^2\, {\int_0^\Delta \d\Pg\, W(\Delta, \Pg)\, 
  (\d\Gamma_{77}/\d\Pg) \over \int_0^\Delta \d\Pg\, W'(\Delta, \Pg)\,
  (\d\Gamma_{77}/\d\Pg)} \simeq (4.27\,\GeV)^2\,, 
\eeq
quite consistently with the physical value of $\mbbar(m_b)$.  In comparison,
Eq.~(\ref{dum}) with the NLL result for $W$ and $W'$ gives $(4.50\,\GeV)^2$.  
We learn that if we keep all two-loop corrections, the physical result is quite
independent of whether we calculate it in terms of $W$ or $W'$, while the same
cannot be said about the NLL resummation result.

The measured $B\to X_s\gamma$ photon spectrum together with $W(\Delta, \Pg)$
given by the sum of Eqs.~(\ref{W2loop}) and (\ref{W1m}) determines
$\int_0^\Delta \d\Pg\, W(\Delta, \Pg)\, (\d\Gamma_s/\d\Pg)$, which in turn
determines $|V_{ub}|$ from a measurement of the partially integrated \Pp\
spectrum in $B\to X_u\ell\bar\nu$ using Eq.~(\ref{Wdef}).  The theoretical
uncertainty of $|V_{ub}|$ from perturbation theory alone using this method is
half the error of the results in Table~I.  However, because of the poor behavior
of the perturbation series, the full two-loop calculation of $W(\Delta, \Pg)$
would be desirable.  Furthermore, the perturbative series is likely to improve
if the unknown matrix elements at ${\cal O}(\lqcd/m_b)$ are expressed in terms
of physical quantities, so that the leading renormalon ambiguity in $W$ is
cancelled.  In addition, and probably more importantly, effects of operators
other than $O_7$ need to be included, in particular that of $O_2$ and $O_8$ may
be important.  In the NLL resummed result, including these is straightforward
using Eq.~(4) in \cite{Neubert:2004dd}.  However, for the BLM result one needs
to combine the analytically known virtual contributions~\cite{Bieri:2003ue} with
the bremsstrahlung contributions~\cite{Ligeti:1999ea}, which are only known
numerically for the $O_2$ operator.  Work in this direction is in progress.

In summary, we calculated the order $\alpha_s^2\beta_0$ corrections to the \Pp\
spectrum in $B\to X_u\ell\bar\nu$ decay and studied the uncertainties in
extracting $|V_{ub}|$ using a measurement of the \Pg\ spectrum in $B\to
X_s\gamma$.  We showed that the factor of $m_b^2/\mbbar^2$ in $W(\Delta, \Pg)$
at lowest order naturally becomes $m_B^2/\mbbar^2$ when subleading effects are
included, and results in much reduced perturbative corrections.  We found that
the NLL RGE resummation is of limited use, because the logs that it sums do not
dominate over the non-logarithmic terms.  This may have implications for the
phenomenological usefulness of other applications of RGE resummations in
inclusive heavy to light decays.

\begin{acknowledgments}

We thank Adam Leibovich and Iain Stewart for helpful conversations.
Z.L.\ and M.L.\ thank the Max Planck Institute for Physics
(Werner-Heisenberg-Institute) for its hospitality while some of this work was
completed. 
This work was supported in part by the Director, Office of Science, Office of 
High Energy and Nuclear Physics, Division of High Energy Physics, of the U.S.\
Department of Energy under Contract DE-AC03-76SF00098 and by a DOE Outstanding
Junior Investigator award (Z.L.); and  by the Natural Sciences and Engineering
Research Council of Canada~(M.L.).

\end{acknowledgments}

\end{document}